\documentclass[fleqn,usenatbib]{mnras}

\usepackage{newtxtext,newtxmath}

\usepackage[T1]{fontenc}
\usepackage[dvipsnames]{xcolor}
\DeclareRobustCommand{\VAN}[3]{#2}
\let\VANthebibliography\thebibliography
\def\thebibliography{\DeclareRobustCommand{\VAN}[3]{##3}\VANthebibliography}

\def\new#1{{\bf #1}}

\usepackage{graphicx}	
\usepackage{amsmath}	
\usepackage{multirow}
\usepackage{pdflscape}
\usepackage{xcolor}
\usepackage{ulem}
\usepackage{soul}

\title[Yang et al. Gaia EDR3 Fornax]{An extended stellar halo discovered in the Fornax dwarf spheroidal using Gaia EDR3 }

\author[ Y.B. Yang et al.]{
Yanbin Yang,$^{1}$\thanks{E-mail: yanbin.yang@obspm.fr,~francois.hammer@obspm.fr, yongjun.jiao@obspm.fr, mpawlowski@aip.de}
Fran{\c c}ois Hammer,$^{1}$
Yongjun Jiao $^{1}$
and Marcel S. Pawlowski$^{2}$
\\
$^{1}$GEPI, Observatoire de Paris, Universit\'e PSL, CNRS, Place Jules Janssen 92195, Meudon, France\\
$^{2}$Leibniz-Institut f\"ur Astrophysik Potsdam (AIP), An der Sternwarte 16, D-14482 Potsdam Germany
}

\date{Accepted 2022 March 1. Received 2021 June 24;}

\pubyear{2021}

\begin{document}

\label{firstpage}
\pagerange{\pageref{firstpage}--\pageref{lastpage}}
\maketitle

\begin{abstract}
We have studied the extent of the Red Giant Branch stellar population in the Fornax dwarf spheroidal galaxy using the spatially extended and homogeneous data set from Gaia EDR3. Our preselection of stars belonging to Fornax is based on their proper motions, parallaxes and color-magnitude diagram. The latter criteria provide a Fornax star sample, which we further restrict by color and magnitude to eliminate contaminations due to either Milky Way stars or QSOs. The precision of the data has been sufficient to reach extremely small contaminations (0.02 to 0.3\%),  allowing us to reach to a background level 12 magnitudes deeper than the central surface brightness of Fornax. We discover a break in the density profile, which reveals the presence of an additional component that extents 2.1 degree in radius, i.e. 5.4 kpc, and almost seven times the half-light radius of Fornax. The extended new component represents 10\% of the stellar mass of Fornax, and behaves like an extended halo. The absence of tidally elongated features at such an unprecedented depth (equivalent to \new{$37.94\pm0.16$}~mag ${\rm arcsec}^{-2}$ in V-band) rules out a possible role of tidal stripping. We suggest instead that Fornax is likely at first infall,  and has lost its gas very recently, which consequently leads to a lack of gravity implying that residual stars have spherically expanded to form the newly discovered stellar halo of Fornax.   
\end{abstract}




\section{Introduction}
\label{sec:intro}
The Fornax dwarf spheroidal galaxy (dSph) is the second brightest dSph after Sagittarius in the Milky Way (MW) halo \citep{McConnachie2012}. It has been  discovered using photographic plates \citep{Shapley1938a,Shapley1938b}. Fornax, together with Leo I, Sculptor, Leo II, Sextans, Carina, Ursa Minor (UMi), Sagittarius (Sgr), and Draco, are known as the classical dSphs, since they are relatively bright \citep{Walker2012,Simon2019}. They are small and faint with very low stellar density, 0.0079 $L_\odot {\rm pc}^{-3}$ \cite[about 20 times less dense than that of solar neighborhood, see][]{Hammer2019}, while their member stars show hot kinematics, i.e., large line-of-sight (\textsc{los}) velocity dispersion ($\sigma_{\textsc{los}}$), at an order of 5 to 10 km/s, \citep[e.g.,][]{Walker2009}.  This may indicate large dynamical mass (i.e., dark-matter, DM) compare to their stellar mass, if they are dispersion-supported systems as speculated from their spheroidal appearance.
To understand their nature, a large number of studies have focused on the stellar structures of these dwarf galaxies to seek evidence for MW tides, tidal tails or extra-tidal debris outside the King tidal radius that is obtained by fitting the stellar density profile of a dSph. Except for Sagittarius which is an obvious case that is undergoing tidal stripping by the MW \citep{Majewski2003}, for three of other classical dSphs evidence or hints of tidal-debris were found: Leo I, Carina, and UMi \citep{Sohn2007,Munoz2006,Munoz2005,Palma2003}. Using deep photometry, \citet{Battaglia2012,Battaglia2013} argued that Carina could be the best candidate with tidal tail detection, but this is not confirmed by \citet{McMonigal2014}. The main reason for different conclusions could be the uncertainty of background determination.

In more than two-third (7/9) of the classical dSphs, a "break" has been detected in their surface density profiles. 
These are
Sculptor \citep{Westfall2006},
Carina \citep{Kuhn1996,Majewski2000,Majewski2005,Munoz2006},
Leo~I \citep{Sohn2007},
Draco \citep{Wilkinson2004},
Sextans \citep{Gould1992},
UMi \citep{Kocevski2000,Matinez2001,Palma2003,Munoz2005}
and, of cause, Sgr \citep{Ibata2001,Majewski2003}. 
This break corresponds to an excess of density that extends outside the King tidal radius. It has been argued \citep[][and references therein]{Westfall2006} to be evidence that MW tides have significantly affected the structure of dSphs. Note that the presence of a "break" is not always acknowledged. For example, \citet{Wilkinson2004} found an obvious break in the density profile of Draco, while \citet{Segall2007} pictured Draco as a featureless dSph, without a break in its density profile.  
In fact, with a careful inspection of the results by \citet[][see their Fig. 10]{Segall2007}, one could notice a break-like feature at the same radius (25 arcmin) as that found by \citet{Wilkinson2004}. The break feature at 25 arcmin is unlikely due to an artifact of background subtraction, since at this radius the density ($\approx 1-2$~stars/arcmin$^2$) is almost an order of magnitude higher than the background ($\approx 0.1$~stars/arcmin$^2$) in both studies. One should note, however, that an over-subtraction of background may lead to an artificial cut-off of extended features. 
The strong contradiction between the conclusions of \citet{Wilkinson2004} and \citet{Segall2007} could be due to the fact that outskirt density profiles of dSphs are very sensitive to background determination, as stressed by \citet{Irwin1995}.

Simulations were designed to test the effects of MW tides on dSphs. They reveal that MW tides may create significant morphological perturbations and breaks in the density profiles \citep[e.g.,][]{Read2006,Munoz2008,Penarrubia2008,Klimentowski2009,Kazantzidis2011}, but cannot inflate the central velocity dispersions to the observed values \citep{Piatek1995,Oh1995}. In order to reproduce the large $\sigma_{\textsc{los}}$, one has to assume large amounts of dark matter to dominate the gravity of dSphs, providing sufficient large velocity dispersions \citep[e.g.,][]{Munoz2008}. These simulations also  support the idea that if MW tides affect the outskirts of these classical dSphs, their cores can be safely approximated by the assumption of dynamical equilibrium \citep{Simon2019,Battaglia2013,Walker2012}. 
Based on this assumption, dynamical analysis using the Jeans equation have been applied to infer the total mass of dSphs via their $\sigma_{\textsc{los}}$ \citep[e.g.,][]{Walker2012,Wolf2010} or via the radial profile of $\sigma_{\textsc{los}}$ for deriving the DM distribution \citep{Walker2011}. If the large $\sigma_{\textsc{los}}$ values are due to  stars in dSphs at dynamical equilibrium with DM \citep{Walker2012}, the assumed DM pushes the theoretical tidal radius well beyond the stellar extent of the dSphs.
Consequently, a DM halo "shields" its stars, including those apparent extra-tidal stars, from any external tides. In such a situation, the "break" in density profiles of dSphs could be interpreted as a second stellar component of the dwarfs, though its origin is still unclear.
Nevertheless, a consensus has been reached that a huge amount of dark-matter (DM) is required to consistently explain the large $\sigma_{\textsc{los}}$, as well as the complex star formation histories of dSphs \citep{Battaglia2013}. A direct confirmation is expected via the annihilation of DM particles \citep{Walker2012}, but no such confirmation has been reported yet.

MW dSphs have been widely recognized as the most DM dominated objects in the universe. This emphasizes their important role in the cosmological context, since it supports a prediction by Lambda cold dark matter ($\Lambda$CDM) cosmology at the smallest galactic scales, despite some debates on $\Lambda$CDM predictions, such as The Plane of Satellites Problem \citep{Pawlowski2018}, the Missing Satellites Problem, the Core-Cusp Problem (specifically on Fornax), and the Too-Big-to-Fail problem \citep{Bullock2017}. 

\begin{table*}
\small
\caption{Observations and methods for studying Fornax.}
\label{tab:method}
{\centering \small
\begin{tabular}{lccccl}
\hline
\hline
  & FOV (sq. deg.) & seeing (") & Bands(limiting mag) & $\Delta$mag$^{\rm a}$ & Method \\ \hline
\citet{Munoz2018}  & $\approx$0.6 &  $<1$ & $g$(25.6); $r$(25.3) & 6.2 & CMD	\\
\citet{delPino2015}  & $\approx$2 & --- & B(23);V & $\approx$5 & CMD	\\
\citet{Coleman2005}    & $\approx$2 & 1.8 &  V(20.7); I & 2.5 & CMD \\
\citet{Battaglia2006}  & $\approx$2.6 & $<1$ & V(23); I(22) & 4.5 & CMD 	\\
\citet{Wang2019DES}    & 25 &$ \approx$1 & $g$(23.5); $r$(23.5)& 4.3 & Matched filter on CMDs\\
\citet{Bate2015}       & 25 & $<1$& $g$(23.1); $r$(22.4); $i$(21.4)& 5 & Matched filter on CMDs\\
\citet{Irwin1995}      & 25 & 2-3 & B(22); R(20-21)& --- & direct counting (mixing star and galaxy, and color)\\
This work & 400  & 0.7$^{\rm b}$ & G (20.8), RP, BP & 3 & CMDs + PM + Parallax\\
\hline
\end{tabular}
}\\
$^{\rm a}$ Estimate of the magnitude range of RGB population covered by observation.
$^{\rm b}$ Note that the 0.7" is the effective angular resolution of Gaia EDR3 \citep{Fabricius2021}. 
\end{table*}

Previously, the success of the DM-dominated scenario for dSph was also strengthened by the 'failure' of DM-free scenarios \citep{Klessen2003,Piatek1995}, known as 'tidal scenario' that were proposed at relatively early time, such as \citet{Kuhn1989,Kuhn1993,Kroupa1997,Fleck2003,Metz2007}. 
All these early propositions of DM-free scenarios assumed that dSphs are long-lived MW satellites since up to $\sim$8 to 10 Gyr ago. Such investigations, especially simulations, were done assuming a single old stellar component. With better resolved stellar populations, it has been revealed that Carina, Fornax, Leo I, UMi and LeoII have their star formation extend until recent 1 or 2 Gyrs \citep{Weisz2014,Pace2020}. In Fornax, a 100-Myr-old stellar populations has been discovered \citep{deBoer2013}, together with an HI cloud that is suspected to be associated with Fornax (Bouchard et al. 2006). 
These facts suggest that their gas may have played an important role at the last stage evolution of dSph dynamics.
The impact of the gas component has been investigated by \citet{Yang2014} for the DM-free case, and by \citet{Mayer2001} and \cite{Mayer2010} for the DM-dominated case.

A recent work \citep{Hammer2019,Hammer2018} identified a strong anti-correlation between the internal acceleration ($\sigma_{\textsc{los}}^2$/$r_{\rm half}$) of dSphs and their distances to the Galactic Centre. Such an anti-correlation is unexpected in the DM scenario, since in such a case the internal acceleration is the DM gravity, for which no specific relation is expected with the Galactocentric distance.  Such an anti-correlation is, however, predicted if dwarfs are experiencing MW tidal shocks in the framework of the impulse approximation for fast encounters \citep{Binney2008}. This scenario requires dSph progenitors to be gas-rich dwarf galaxies approaching the  MW in the past few billion years. In such a case, after loosing their gas due to ram pressure induced by the MW halo gas, dSph stars spherically expand because of the loss of a major part of the dSph potential. Due to the spherical expansion, part of the residual stellar body (about 25\%, see \citealt{Hammer2019}) exchanges kinetic energy through MW tidal shocks, which may explain the large velocity dispersions of dSphs without the need for DM. This mechanism has been previously investigated with numerical simulations by \citet{Yang2014}.  It predicts anisotropic velocity dispersions, with large values along the line of sight, which is nearly the Galactic Center direction, and small values on the axis perpendicular to the dSph orbital motion \citep{Hammer2018}. 
Possibly, this is another mechanism that could explain the nature of dSphs, without invoking a dominant dark matter component.
Gaia\footnote{https://www.cosmos.esa.int/web/gaia/home} DR2 (and more recently, EDR3) reveal dSph orbital motions that appear not consistent with that of DM dominated MW satellites (or sub-haloes). This is because they lie too close to their pericenters \citep{Hammer2020,Li2021}, they are preferentially distributed into a Vast Polar Structure \citep{Pawlowski2014} within which they orbit, and their tangential velocities and angular momenta are excessively large, suggesting a recent infall \citep{Hammer2021}.

Gaia will bring a complete proper motion (PM) coverage of the whole sky, which provides an independent method (comparing to the widely used color-magnitude diagram, i.e., CMD method) to distinguish stars belonging to dwarf galaxies from that of MW because they move differently. This will be helpful to reach a deeper and more accurate background determination, allowing to further identify possible tidal debris around dSph and weak structures in dSph. 
In any scenario, MW tides may play a role in shaping dSphs, but how and by how much? 
Answering this fundamental question requires to reach the true edges of dSphs by eliminating most contamination from non-member stars. Besides observational depth, the requirement of a pure member sample is preponderant, i.e., removing the contamination from MW stars and background Quasi-Stellar Objects (QSOs). Member stars of dSphs in the RGB branch are bright enough to be observed by Gaia, and so one should be able to robustly retrieve the dSphs' true extent using a highly pure member sample. 

In this paper, we present a case study on Fornax, using Gaia EDR3 published in December 2020 \citep{Brown2020}. 
Thanks to its homogeneous coverage and data quality, we can explore the data over a very large area.
Both coverage and calibrations across large field are difficulties for ground-based and mosaic-type observations on dSphs.

\section{Methodology}
\label{sec:method}
Our goal is to search possible faint structures in the outskirts of Fornax using Gaia data.
This is a challenging task because Gaia data is limited to a relative shallow depth, i.e., $G<21$~mag, covering about a 3 magnitude range of the RGB branch of Fornax\citep{Helmi2018}. 
In Table~\ref{tab:method}, we compare major photometric works that have been dedicated to the study of Fornax' structure. \citet{Munoz2018} carried out the deepest observation, but only focused on the central region. \citet{delPino2015}, \citet{Coleman2005} and \citet{Battaglia2006} carried out observations that barely cover the nominal tidal radius of Fornax. Thus, the estimation of background in these studies may have more uncertainties, as acknowledged in Battaglia et al.. \citet{Wang2019DES}, \citet{Bate2015} and \citet{Irwin1995} presented deep photometry studies covering the wider field around Fornax. All these studies, except \citet{Irwin1995}, used CMDs to pre-select member candidates of different stellar population such as RGB, red clump, etc.. In doing so, many Galactic foreground stars and compact background galaxies that differ in color from Fornax stellar populations can be excluded from the candidate sample. All studies performed star/galaxy separation and exclude extended sources before applying the CMD selection, except \citet{Irwin1995} (due to poor seeing conditions).

When using the CMD method to select member candidates, there are always non-member sources selected into the candidate sample due to their colors blending with Fornax's stellar population. The fraction of such blending is partly intrinsic and partly dependent on photometric accuracy: the better the accuracy, the lower the contamination. These contaminating sources will appear as a background in the spatial distribution which can be removed statistically. Theoretically, such a background due to contaminating sources should be smoothly distributed compared to the structure of Fornax. Even if the background is not flat, it could be modelled by fitting a surface if necessary, and the fluctuation of the background will be recognized as an uncertainty. Importantly, the uncertainty of such a background directly defines the limit to which we are able to probe faint structures of, and around, a dSph such as Fornax.

Gaia data reach more shallow photometry and less accurate color measurement compared to the ground-based observations (see Sect.~\ref{sec:data}). Yet, the Fornax RGB branch can still be recognized easily. We may select only RGB candidates to study its morphology and possible debris under the assumption that the RGB is representative of the stellar population of the galaxy, knowing that Fornax is dominated by an old stellar population even though recent star formation has been discovered \citep{Weisz2014,deBoer2013,delPino2013,Rusakov2021}.

If there were tidal tails in Fornax, detecting them requires that we search as far as possible from the object, and that we reach the real background where there are no member stars by definition. Thus, choosing a secure reference as our background region is mandatory. A symmetric region, and as far as possible around the object in question, would be ideal because possible tidal debris are expected to distribute along a specific direction. 
If member stars of Fornax were to fall into the background region, for example due to the presence of a long tidal tail, the background will be overestimated in principal, which may lead to artificial truncation of the density profile and will reduce or even remove possible signatures of faint structures. Gaia's full-sky coverage easily provides us the opportunity to perform a study over a large area, which will provide us with sufficient counts in the background region to accurately evaluate the uncertainty of the background level.

Table~\ref{tab:method} compares the seeings from all studies. In term of spatial resolution, Gaia EDR3 is the best amongst all studies in the table. A good effective spatial resolution helps to obtain a better completeness of the source catalog, and hence reduces the effect of crowding in dense region. The crowding effect could reduce the estimated spatial density in the central region, so that the measurement of the shape of an object, such as its ellipticity and radius, will be affected. As shown by \citet{Bate2015}, the crowding effect may cause an overestimation of the Sersic radius by about 5\%. This could be one of the reasons\footnote{On the other hand, an over-subtraction of background could also cause an overestimation of the Sersic radius.} why other studies e.g., \citet{Battaglia2006} or \citet{Munoz2018}, found relatively larger Sersic radii without considering the effect of crowding. 
\citet{Irwin1995} used observations with worse seeing conditions, 
which should lead to a more severe crowding effect. 
To test this, we fit a Sersic profile to their density profile of Fornax (the data in their Table 3). We found a Sersic radius of 21.9 arcmin (see Appendix~\ref{sec:ih95}), which is the largest amongst all studies of Fornax, and consistent with our analysis of the impact of seeing on the study of the dwarf's structure. The other aspect, shallow photometry, also helps to reduce the crowding effect. Bright or faint RGB stars are the same stellar population, thus, they should trace the same structure of Fornax. However faint RGB stars are numerous, hence could result in a distribution affected by crowding, especially towards the center of the galaxy.  Under such consideration, a shallower depth of photometry may be a better way to probe the structure in the central region.  
A drawback with shallow photometric data will be less significance in statistics. Nevertheless, our major interest is to search faint structure in the outskirts where the crowding effect becomes negligible if using Gaia data. 

The main mission of Gaia is to provide PM measurement of all sources, which is an independent parameter space from the CMD.
As Fornax is moving around the MW, in principle we should be able to distinguish Fornax member stars from non-member sources in a certain region in PM space. Thus, we could select member stars by PM. 
Similar to the problem with the CMD-selection method, 
there will be contaminating sources selected as candidates due to their intrinsic PM overlapping with that of Fornax, and the error bars in PM will introduce more contamination.

By combining the selections from both CMD and PM, i.e., two independent parameter spaces, principally we are able to eliminate more contaminating sources, and hence reducing the level of the background in the spatial distribution.  In addition, parallaxes could be used for eliminating foreground stars close the Sun in case they fall in the RGB and PM selections.

Fornax is located at high Galactic latitude ($b=-65.6$ deg) where the foreground contamination from the MW is relatively small when compared to many other dSphs at lower galactic latitude. We may thus expect to reach deeper background depth. 
Finally, another tricky problem is that we have to keep contaminating sources in sufficient numbers to accurately measure the background level, and especially its uncertainty, in order to characterise the depth and the significance of faint structures that we are seeking. In other words, it is not necessary to completely remove contamination, but we need to optimize the selection with all the above considerations to obtain the least contaminated candidate sample and to keep sufficient counts to evaluate the spatial background level.

Note that we do not consider the method of likelihood selection, e.g., \citet{Pace2019} where a prior spatial distribution of a Plummer sphere is assumed for the target. Such an assumption may lead to biases if we want to search faint and unknown structures. In brief, in term of methodology of member candidate selection, we will use Gaia EDR3 data to select member candidates having the color of Fornax' RGB, having similar PM as Fornax' mean motion, and being reasonable distant from the Sun. All selection conditions will be scaled by the error bar of each relevant quantity.

\section{Data}
\label{sec:data}
We have chosen a field of 20 by 20 degrees centered on Fornax. This very large area, more than 30 times larger than Fornax in diameter (half light diameter of 0.6 degree, Munoz et al. 2018), should be enough to reach and robustly determine the outskirts of Fornax. First, we selected a raw sample ($S_{\rm raw}$) of sources from EDR3\footnote{Gaia Archive at https://gea.esac.esa.int/archive/} with the following conditions to control the quality of the data:  
Not \texttt{duplicated\_source}; 
Not QSO;
with color \texttt{bp\_rp} measured;
with astrometry solutions (either 5-parameters or 6-parameters);
\texttt{G~<~20.8};
\texttt{ruwe < 1.4};
and \texttt{$C^{*} < 1.0$}. The $C^{*}$ is the corrected \texttt{phot\_bp\_rp\_excess\_factor} introduced by \citet{Riello2020}, which is very helpful to remove background galaxies (see their Figure~$\!$21); bright QSOs can be identified by cross-matching with the confirmed QSO sample provided in the Gaia EDR3 database, i.e., the table \texttt{gaiaedr3.agn\_cross\_id}. 
For this raw sample, we have median uncertainties in photometry: $0.008$ in \texttt{G}-band, $0.15$ in \texttt{BP} and $0.09$ in \texttt{RP}, resulting in a less accurate in color (\texttt{BP}$-$\texttt{RP}) than the ground-based observation, e.g., \citet{Battaglia2006}. Nevertheless, the RGB branch of Fornax stars can still be recognized unambiguously. 
We checked Galactic extinction over the field, which is  $E(B-V)= 0.04$ magnitude on average, and then we do not apply any correction in the following analysis. For a better interpretation of internal structures in both morphology and kinematics, we have adopted the projection introduced by \citet{Helmi2018,Luri2020}, including the corresponding transformation of proper motions:
\begin{equation} \label{eq:1}
\small
\begin{split}
x & = \cos{\delta} \sin{(\alpha - \alpha_{\textsc c})} \\
y & = \sin{\delta} \cos{\delta_{\textsc c}}  - \cos{\delta} \sin{\delta_{\textsc c}}\cos{(\alpha - \alpha_{\textsc c})} \\
\begin{bmatrix} 
\mu_{x}  \\
\mu_{y} 
\end{bmatrix} & =  \mathsf{M} 
\begin{bmatrix} 
\mu_{\alpha*}  \\
\mu_{\delta} 
\end{bmatrix} \\
\mathsf{C}_{\mu_{x,y}} & = \mathsf{M}\, \mathsf{C}_{\mu_{\alpha*,\delta}} \mathsf{M}^{T}, \\{\rm where}\\
\mathsf{M} & = \!\!
\begin{bmatrix} 
\cos{(\alpha - \alpha_{\textsc c})}  &  -\sin{\delta} \sin{(\alpha - \alpha_{\textsc c})}  \\
\sin{\delta_{\textsc c}} \sin{(\alpha - \alpha_{\textsc c})} & \!\!\! \cos{\delta}\cos{\delta_{\textsc c} 
\!+\!\sin{\delta} \sin{\delta_{\textsc c}} \cos{(\alpha - \alpha_{\textsc c})}
} 
\end{bmatrix},
\end{split}
\end{equation}
$\mathsf{C}_{\mu_{x,y}}$ and $\mathsf{C}_{\mu_{\alpha*,\delta}}$ are covariance matrix. 
At the morphological center we have $({\mu}_x,{\mu}_y)_\textsc{c} = ({\mu}_{\alpha*},{\mu}_\delta)_\textsc{c}$. All the following analysis steps are carried out in this projection frame, unless specified otherwise.

\begin{figure}
\includegraphics[width=8.5cm]{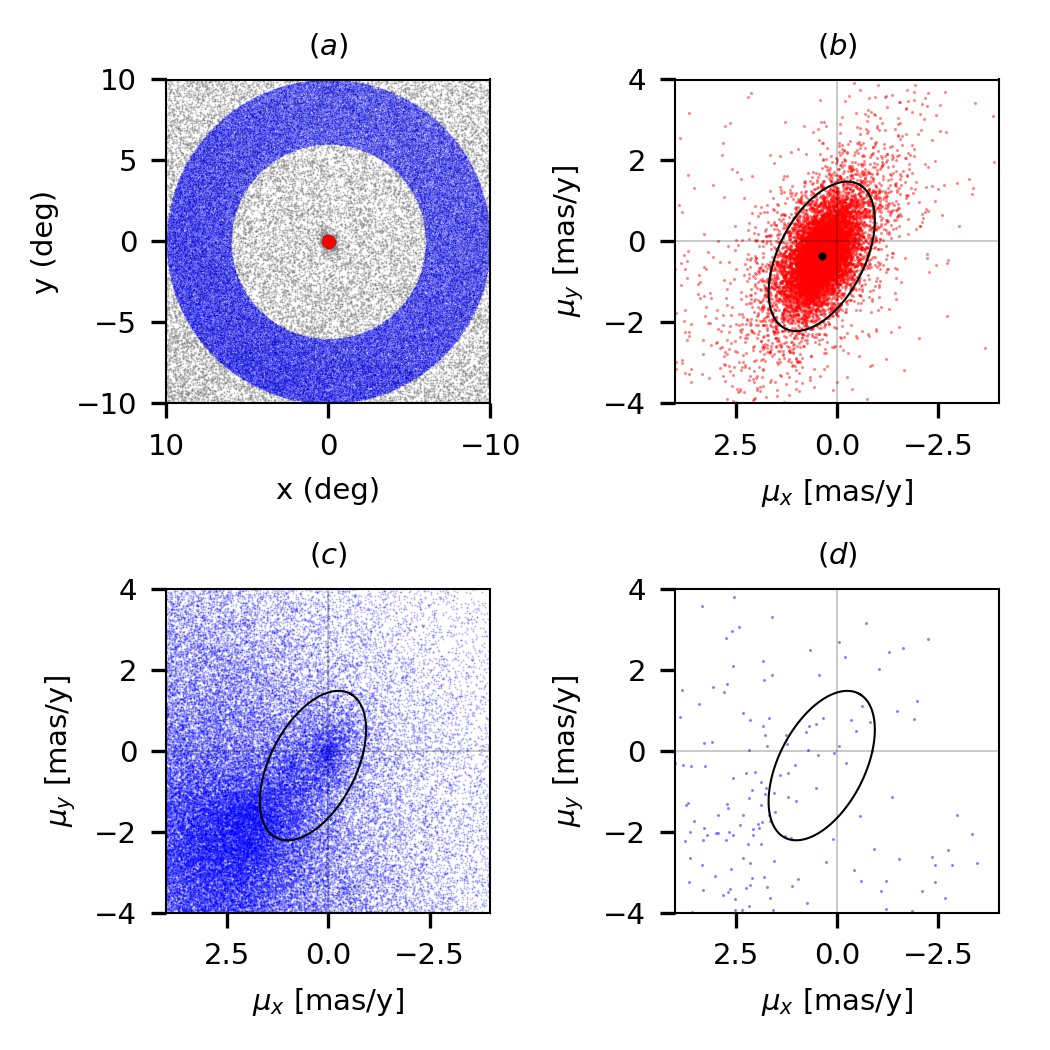}
    \caption{Panel~($a$) spatial distribution of sources showing the spatial definition of the object sample, $S_{\rm obj}$ (red, a circle of 0.4-degree radius), the reference sample, $S_{\rm ref}$ (blue, an annulus of 6-10 degree), and 10\% of the raw sample, $S_{\rm raw}$ in gray. Panel~($b$), PM distribution of $S_{\rm obj}$ showing that Fornax member stars concentrate in an elliptical region. The average PM of Fornax is marked by the black dot. The ellipse, also in panel ($c,d$), has a major axis of 2.0 mas/yr, a minor axis $1.06$~mas/yr and an orientation 62.9 degree (measured from $x$-axis, anti-clockwisely). Panel~($c$), PM distribution for $S_{\rm ref}$, where we can see a concentration of sources at zero proper motion, which could be mostly due to background compact galaxies, such as QSOs; and that MW stars distribute very broadly peaking around $(\mu_x,\mu_y) = (2.5,-2)$ and overlapping with Fornax member stars. 
    Panel~($d$) illustrates the distribution of a comparison sample, i.e., possible contaminating sources, selected from $S_{\rm ref}$ in an area that corresponds to that of $S_{\rm obj}$, which means both of the comparison sample and $S_{\rm obj}$ have the same sky coverage.
Hence, the source distribution in panel $b$ and $d$ are directly comparable. From this comparison, one could infer the level of contamination. This comparison sample of contamination is taken arbitrarily from $S_{\rm ref}$ as a ring from 7.0 to 7.0114 degree.
}
    \label{fig:check}
\end{figure}

\begin{figure}
\includegraphics[width=8.5cm]{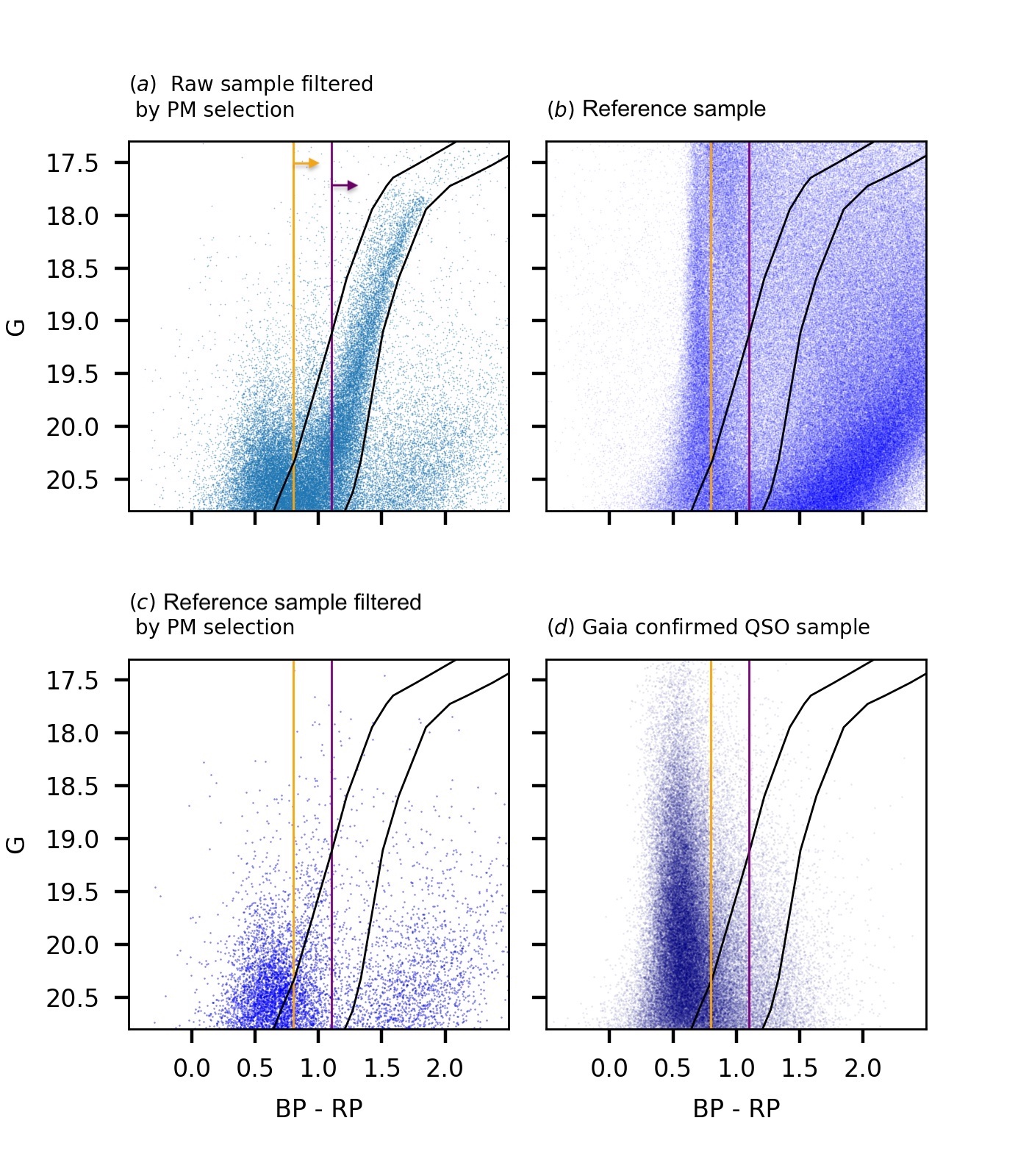}
    \caption{CMD for different samples, i.e., the title of each panel where "PM selection" refers to the selection method in Sect~\ref{sec:selPM}.  In all panels, solid black lines set limits for the selection of RGB stars in each CMD; vertical orange and red lines illustrate color cuts at \texttt{BP-RP}=0.8 and 1.1, respectively. }
    \label{fig:cmdselections}
\end{figure}

\begin{figure*}
\includegraphics[width=18cm]{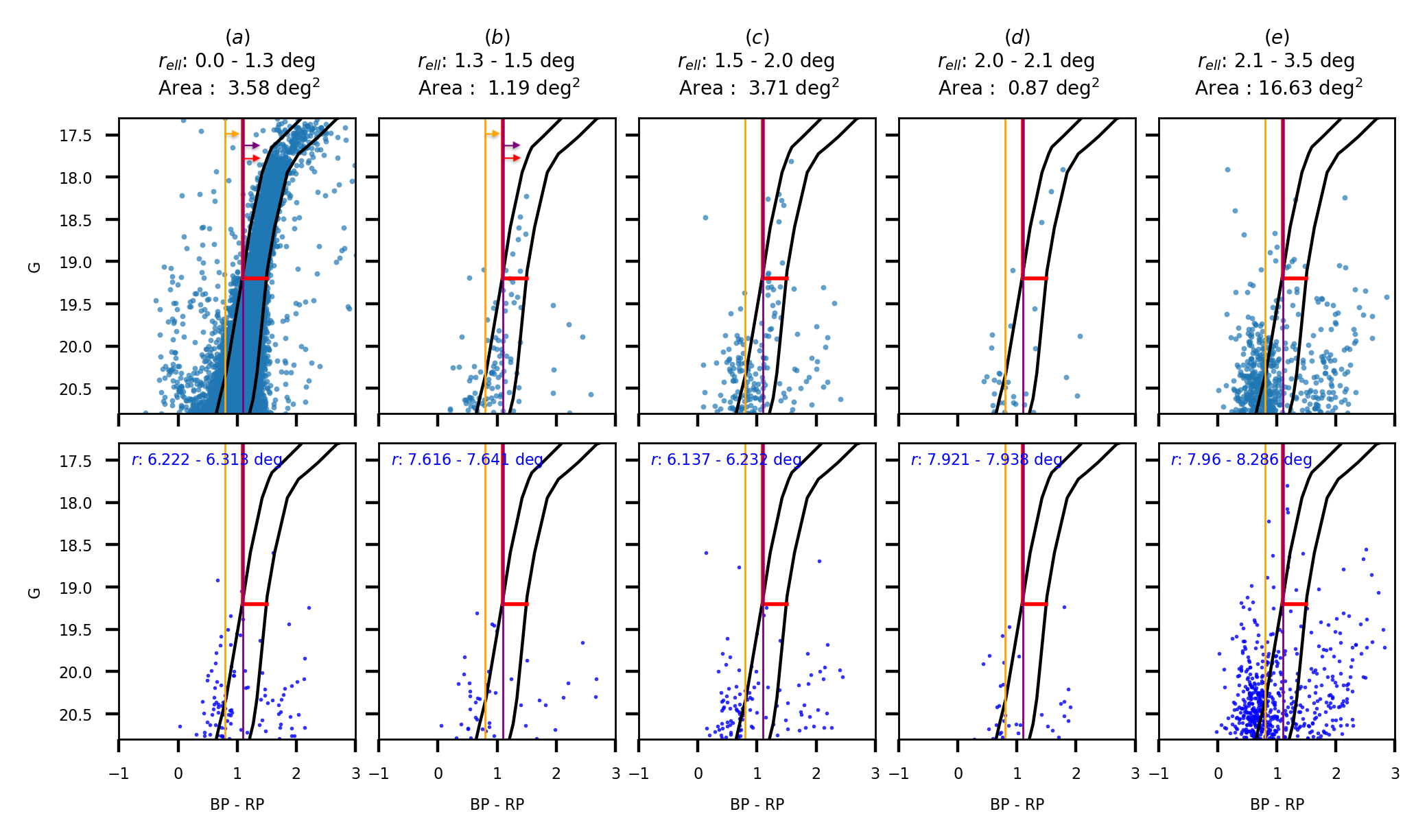}
    \caption{Top row: CMDs for the sample $S_{\textsc{\,pm}}$, as a function of elliptical annuli. The range of each annulus and the corresponding area are indicated at the top of each panel. The vertical orange and purple lines indicate color cut at \texttt{BP-RP} larger than 0.8 and 1.1, respectively. The red line marks \texttt{BP-RP}=1.1 and \texttt{G} = 19.2, the conditions for defining $S_2$.  Bottom row: CMDs of  possible contaminating sources that corresponds to the same sky coverage for each elliptical annulus in the top row. Each sample in the bottom row is taken randomly from $S_{\rm ref}$ in a ring, which radius interval is indicated at the top of the panels. } 
    \label{fig:cmdofpm}
\end{figure*}

\section{Member star candidates}
\label{sample}

In order to detect possible faint structures far from the center of Fornax, we use the combination of several available parameters, such as the color-magnitude diagram (CMD), proper motions (PMs), and parallaxes to remove at best contaminations from QSOs or Galactic stars. For each parameters we carefully define specific procedures for selecting stars associated to Fornax. Final samples are optimized and carried out iteratively in all parameter spaces in order to obtain the best signal-to-noise ratio (S/N) for the surface density profile.

\subsection{Selection by PM} 
\label{sec:selPM}
As indicated in Figure~\ref{fig:check}-$a$, we defined an object sample ($S_{\rm obj}$, red) of sources that are located within the central 0.4-degree radius and a reference sample ($S_{\rm ref}$, blue) based on a 6-10 degrees annulus. The latter is also defined as the reference or background region throughout of the paper.
Fornax member stars (see panel $b$) concentrate in an elliptical region in PM space. At Fornax' distance 147 kpc, 10 km/s on sky is equivalent to 0.014 mas/yr, which is far smaller than the median uncertainty (0.391~mas/yr) for an individual source in $S_{\rm raw}$. Thus, this distribution in PM space reflects only the uncertainties of the astrometry determination and a global anti-correlation between $\mu_x$ and $\mu_y$ for member stars. We measured the average shape of this distribution, indicated in panel ($b$) by an ellipse. The details of this calculation is explained below. 
MW stars, i.e., $S_{\rm ref}$ (see panels $c$) show quite a scattered distribution with a peak shifted from that of Fornax stars. 
In panel~($d$), we illustrate the level of contaminating sources for $S_{\rm obj}$ by showing a comparison sample, see more details in the figure caption. 

If a star has its PM consistent with the mean PM of Fornax within its error bar, it can be considered as a member candidate. As we mentioned above, the dominant Fornax population shown in Figure~\ref{fig:check}-($b$) actually reflects the random error distribution of PM for member stars. If we define a selection region following this distribution, we will obtain a sample of member candidates. The distribution can be characterized by an ellipse centered on the mean PM of Fornax denoted by $({\mu}_x,{\mu}_y)_\textsc{c}$.
On the other hand, the ellipse should not extend too far from the mean PM of Fornax otherwise the peak distribution of MW star will be included, increasing the contamination. In Figure~\ref{fig:check}-($b$), the size of the ellipse is set at $2.0$~mas/yr in major axis, which seems a best choice according to our following analysis. 
Our goal is to search Fornax member stars in the full field of $S_{\rm raw}$, thus, principally we should not apply any spatial condition for the selection. 
With all these considerations, we implemented the following procedure to perform the selection with PM.

In order to characterise the PM distribution of Fornax' stars as seen in Figure~\ref{fig:check}-($b$), we first prepare a core-sample which is defined within an elliptical radius of 1 degree according to the morphological parameters of Fornax, i.e., the on-sky center ($\alpha_\textsc{c}, \delta_\textsc{c}$), ellipticity and position angle (PA). These parameters can be initialized with values from the literature, e.g., \citep{Battaglia2006}, and revised later using our own selected sample. 
With the core sample, we can calculate the corresponding error-weighted average PM, i.e., $({\mu}_x,{\mu}_y)_\textsc{c}$ and the axis-ratio and orientation of the ellipse representing the PM distribution. 
Note that the properties of an elliptical distribution are calculated by averaging the source counts, following the method in \citet{Bertin1996}.

To define the member candidate sample selected by PM, namely "$S_{\textsc{\,pm}}$", we required the following two conditions:

\begin{enumerate}
   \item The PM of a candidate must follow the correlation between $\mu_x$ and $\mu_y$,  i.e., fall inside the ellipse that characterises the PM distribution, e.g., Figure~\ref{fig:check}-($b$). The size of the ellipse is limited to $2.0$~mas/yr as a global condition.
This cut-off is 5 times larger than the median uncertainty of individual sources, i.e, 0.391 mas/yr, so it is reasonably large to fully over the error distribution of member stars in PM space, while avoiding the increasing contamination beyond the ellipse in Figure~\ref{fig:check}-($c$) . 
  \item As uncertainties of measurements are linked to the magnitude of sources, we applied a more restrict selection by requiring that each source has its PM ($\mu$) to be consistent with $({\mu}_x,{\mu}_y)_\textsc{c}$ within 3 times of its own uncertainty. The uncertainties of $\mu$ for each star is calculated via error propagation by considering the correlation coefficient between $\mu_x$ and $\mu_y$. 
\end{enumerate}

The PM selection procedure is done by successive iterations, because each change of the morphological parameters of Fornax and the characterizations of the PM distribution will affect the candidate sample. In doing so, we kept a full consistency between selection conditions and the corresponding output. This selection by PM is applied, respectively, to the three Fornax-member samples that will be introduced in the next section. All calculated parameters are listed in Table~\ref{tab:fornax}.

\subsection{ Selection by CMDs} 
\label{sec:selCMD}
Figure~\ref{fig:cmdselections}-$a$ shows the CMD of $S_{\textsc{pm}}$ where the RGB branch is seen clearly. We define sources enclosed by two black lines in the CMD as sample $S_{\textsc{\,rgb}}$ that is selected by only colors and magnitude. 
By comparing to panel-$b$ and $c$, we notice that many bright MW stars have been excluded, leaving fainter source around $\texttt{BP-RP}=0.6$ that may contaminate faint Fornax RGB stars when $\texttt{BP-RP}$ less than around 1. Further checking (see panel-$d$) we notice that these contaminating sources may be faint QSOs that are not identified yet. Note that the confirmed QSOs have been excluded when we selected $S_{\rm raw}$.

To avoid contamination, we may consider more or less restrictive conditions, such as using further limits in \texttt{BP-RP} (see vertical lines in Figure~\ref{fig:cmdselections}) and combine them with $S_{\textsc{\,rgb}}$ to obtain CMD selected sample.
To maximize the total number of selected Fornax members, we limit $S_{\textsc{\,rgb}}$ in the magnitude range $17.3 < \texttt{G} < 20.8$, which covers the tip of RGB and guarantees the data quality. Fainter than  $\texttt{G} > 20.8$, the uncertainties in astrometry and photometry increase very quickly, which reduces the efficiency in separating stars of Fornax from those of the MW.
After testing many possible selections by limiting magnitude range and cut-offs in \texttt{BP-RP}, we finally choose three representative selections by CMD to establish the Fornax-member samples (see Sect~\ref{sec:final}).

In the top row of Figure~\ref{fig:cmdofpm}, we plot CMDs as a function of elliptical annuli, which is an efficient way to explore how far from the center Fornax stars could distribute. 
In the bottom row of Figure~\ref{fig:cmdofpm}, we plot, illustratively, comparison samples of possible contamination that correspond to each elliptical annulus. Each comparison sample in the bottom row is chosen arbitrarily from $S_{\rm ref}$ by requiring the same sky coverage as in the corresponding top panel. In doing so, we can compare the CMD distribution directly in number between member candidates (top) and background (bottom).
For example, in panel-$c$, if we count the number of points inside the RGB region for top and bottom panel respectively, and subtract them, we find that there is an excess of counts of around 30 in the top panel.  This is a strong signal that Fornax member stars exist in this annulus. In panel-$d$, we found an excess count of 4, which is a much weaker signal of about two sigma. This suggests that Fornax' member stars may extend out to 2-degree elliptical radius.
Properly subtracting the contamination can only be done statistically in the following density analysis. The illustration in Figure~\ref{fig:cmdofpm} is only to visualize the significance of Fornax' RGB stellar populations at different radii.

\subsection{Selection by parallax}
\label{sec:selparallax}
Fornax is located at a distance $147\pm 4$ kpc from the Sun \citep{deBoer2016}, corresponding to a parallax $0.00680$~mas, which is far smaller than the typical error of individual sources in Gaia data. Hence, in principle the parallax distribution for Fornax stars in EDR3 reflects only the uncertainties in parallax determination. Thus we applied a gentle filter by rejecting some sources with \texttt{|parallax\_over\_error|}~$ > 3.5$, see Figure~\ref{fig:verify}. Correspondingly, we defined $S_{\textsc{plx}}$ for selection \texttt{|parallax\_over\_error|}~$ <= 3.5$ to be combined with other conditions (see Table~\ref{tab:stat}).

\begin{table} 
\small
\caption{Fornax parameters derived from this work.}
\label{tab:fornax}
{\centering \small
\begin{tabular}{lrrrr}
\hline
\hline
Parameters  & S0 & S1 & S2 \\ \hline
$\alpha_\textsc{c}$(J2000.0)  &   $2^\mathrm{h}39^\mathrm{m}50.9^\mathrm{s}$  &  $2^\mathrm{h}39^\mathrm{m}50.9^\mathrm{s}$  & $2^\mathrm{h}39^\mathrm{m}51.0^\mathrm{s}$  \\
$\delta_\textsc{c}$(J2000.0)  &   $-34^\circ30{}^\prime53{}^{\prime\prime}$ & $-34^\circ30{}^\prime54{}^{\prime\prime}$ & $-34^\circ30{}^\prime49{}^{\prime\prime}$  \\
Ellipticiy &  0.325 & 0.317  & 0.321  \\
Position angle (deg)$\!\!\!\!\!\!$  &  47.7 &  47.3  &  46.8  \\
\hline
${\mu}_{\alpha*,\textsc{c}}$(mas/yr) $\,\,\,^\dagger$$\!\!\!\!\!\!$& $0.381\pm 0.001 $  &$0.381\pm 0.001 $&  $0.381\pm 0.002 $  \\
${\mu}_{\delta,\textsc{c}}$(mas/yr)  $\!\!\!\!\!\!$$\!\!\!\!\!\!$ & $-0.367\pm 0.002$  &$\!\!\!\!\!\!$$-0.367\pm 0.002$&  $\!\!\!\!\!\!$$-0.366\pm 0.002$ \\
Axis-ratio &  0.531 & 0.533  & 0.610  \\
Orientation &  62.9 & 64.4 &  65.7 \\
\hline
\end{tabular}
}\\
$^\dagger$For PMs, only the random error is quoted here, while for the systematic errors one may refer to \citet{Li2021} : 0.018 and 0.019 mas/yr, respectively. 
\end{table}

\subsection{Three final samples of Fornax member candidates} 
\label{sec:final}
\begin{figure}
\includegraphics[width=8.5cm]{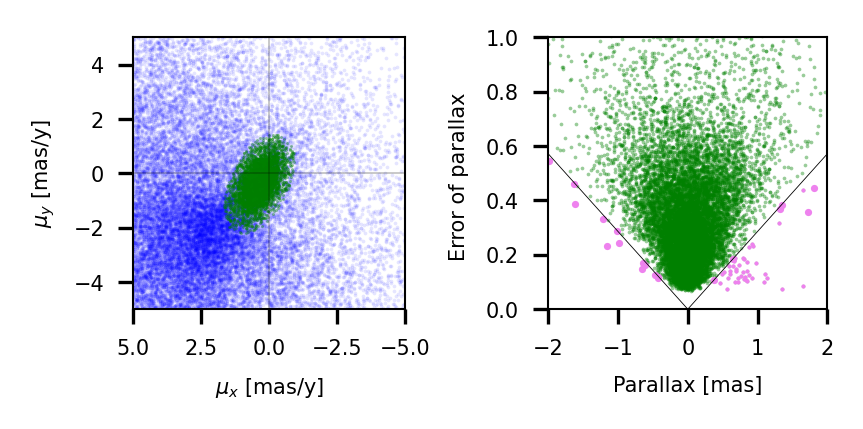}
\includegraphics[width=8.5cm]{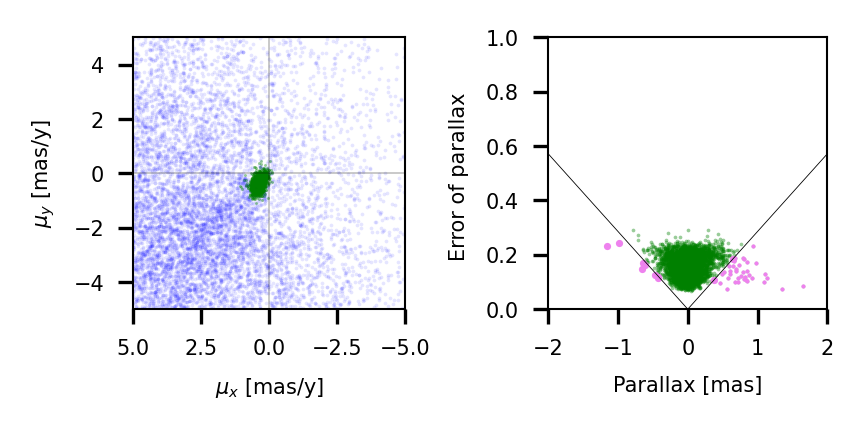}
    \caption{Verification of the final samples (green dots) in PM space (left panels) and parallax versus parallax error (right panels). Only $S_1$ and $S_2$ are shown in the top and bottom row, respectively. $S_0$ has almost identical distribution as $S_1$ in these two parameter spaces, except with a larger number of sources. Blue dots are the corresponding reference sample in the magnitude ranges of $S_1$ and $S_2$, respectively. In the right panels, violet dots are those rejected by the parallax condition indicated by the two black lines  (see more in Sect~\ref{sec:selparallax}), and larger dots are those within the central 1 degree radius.
    }
    \label{fig:verify}
\end{figure}

After applying all the selection conditions we obtain our final samples of Fornax member candidates $S_{\textsc{0}}$ , $S_{\textsc{1}}$, and $S_{\textsc{2}}$ with different conditions, as summarized in Table~\ref{tab:stat}. 
$S_0$ selects the largest number of member candidates as possible, and it provides the best S/N for the following data analysis. $S_1$ is defined by limiting the color to be redder, in order to better remove QSOs and  MW stars that overlap in color with Fornax RGB, as seen in Figure~\ref{fig:cmdselections}. Thus $S_1$ provides a less contaminated background that could still be determined accurately given its large number of stars. $S_2$ is a very strict sample of bright stars having the best measurements in PM and parallax (see Figure~\ref{fig:verify}). With $S_2$, we may probe the extent of Fornax based on bright stars, also to verify the structure analysis from the other two samples. In the following we analyse the three samples in parallel, in order to consolidate stable and variable structures, as well as the robustness in our structure analysis. The contamination fraction (CC) from background counts within 2.1 degree radius is estimated and given in Table.~\ref{tab:stat}.

\begin{figure}
\includegraphics[height=4.4cm]{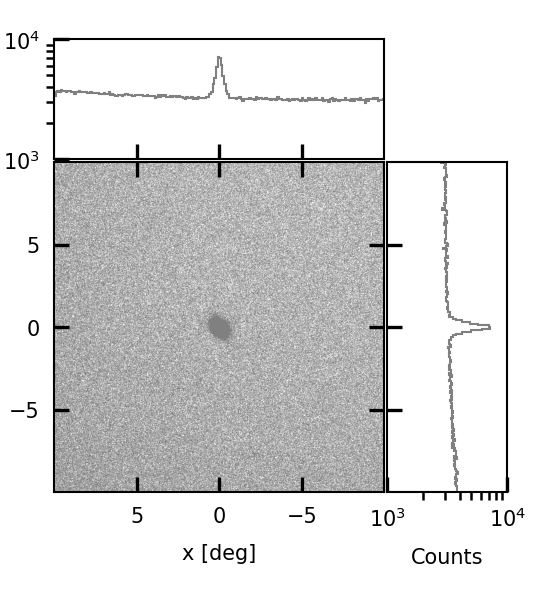}\includegraphics[height=4.4cm]{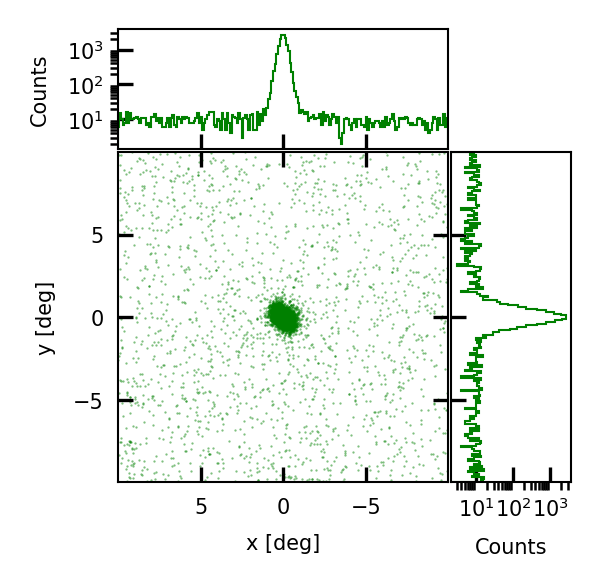}
    \caption{ 
{\it Left panel: } Inspection of background for the raw sample $S_{\rm raw}$. Histograms of star counts are shown respectively for $x$ and $y$ projection. Gradient of counts can be clearly seen in both projections, which is due to the MW stars. 
{\it Right panel :} The same figure as the {\it Left}, but for $S_{0}$, where the gradient of $S_{\rm raw}$ has clearly disappeared, indicating our selection method is very efficient in removing MW stars. 
}
    \label{fig:hist}
\end{figure}

We verified the member sample selection in PM space and parallax space, see Figure~\ref{fig:verify}. The global background is obtained through the 6-10 degree annulus of Figure~\ref{fig:check} (see panel(a)), for which the mean density and Poisson noise can be robustly determined for each of the three Fornax samples (Table~\ref{tab:stat}). 
Figure~\ref{fig:hist} compares the spatial distribution of $S_{\rm raw}$ and one of the Fornax-member sample, $S_{0}$. It shows that the gradient observed in the background counts (left panel) disappears in the right panel, meaning that most of the MW stars have been excluded. Moreover, it illustrates that there is no obvious large scale structure, such as tails in the large field of view.

\begin{table*} 
\small
\caption{Statistic properties of the three Fornax candidate samples.}
\label{tab:stat}
\begin{tabular}{lccccccccc}
\hline
\hline
\multicolumn{2}{c}{{Samples and  Conditions }} &  $N$    &   CC$^{\dagger}$ & $\Sigma^{\rm 2D}_{\rm bg}$  & ${\rm S.B.}_{\rm lim}$ $^{\dagger\dagger}$  &  $\Sigma^{\rm 1D}_{\rm bg}$   & ${\rm S.B.}_{\rm lim}$ $^{\dagger\dagger}$&  (S/N)$_*$  & Cnt.$_*^{\rm net}$  \\ 
    &&&  &  ($10^{-3}$arcmin$^{-2}$)  & mag/arcsec$^2$ & ($10^{-3}$arcmin$^{-2}$)  & mag/arcsec$^2$ & \\  \hline
    &&& $\!\!\!\!\!\!$($r_{\rm ell}<$2.1deg)$\!\!\!\!\!\!$ & \multicolumn{2}{c}{ 2D surface density} & \multicolumn{2}{c}{ 1D surface density} & \multicolumn{2}{c}{($1.3<r_{\rm ell}<2.1$)} \\ 
$S_{\textsc{0}}$:$\!\!\!\!\!\!$ &  $S_{\textsc{base}}$$^{\dagger\dagger\dagger}$  {\scriptsize \&}  ($\texttt{G} < 20.8$ {\scriptsize \&} $\texttt{BP-RP} > 0.8$)   & 21359 &53 / 17140 & 1.63$\pm$1.77 &  33.55 & 1.625$\pm$0.047  & 36.00 & 9.1 & 117 \\  
$S_{\textsc{1}}$:$\!\!\!\!\!\!$ &  $S_{\textsc{base}}$ {\scriptsize \&}  ($\texttt{G} < 20.8$ {\scriptsize \&}  $\texttt{BP-RP} > 1.1$) &  12818 &14 / 11669& 0.41$\pm$0.91 & 33.86 & 0.421$\pm$0.024  & 36.59 & 6.8 & 55 \\
$S_{\textsc{2}}$:$\!\!\!\!\!\!$ &  $S_{\textsc{base}}$ {\scriptsize \&}  ($\texttt{G} < 19.2$ {\scriptsize \&}  $\texttt{BP-RP} > 1.1$) &  3883 &1 / 3769& 0.03$\pm$0.25 &  33.99 & 0.030$\pm$0.006  & 37.94 & 4.4 & 20 \\
\hline
\end{tabular}\\
$^{\dagger}$ Estimated contamination counts from background density over the total counts inside $r_{\rm ell}<2.1$ degree.\\
$^{\dagger\dagger}$ V-band surface brightness limits are defined at the 3-sigma level of above the mean background of each sample; all values have the same error $\pm0.16$ mag.\\
Note that the last two columns are the signal-to-noise ratio and net counts of member stars (after subtracting the corresponding background counts) in between 1.3 and 2.1 elliptical radius.\\
$^{\dagger\dagger\dagger}$ $S_{\textsc{base}}$ = ($S_{\textsc{pm}}$  $\cap$ $S_{\textsc{rgb}}$ $\cap$ $S_{\textsc{plx}}$), which are defined in Sect~\ref{sample}. \\
\end{table*}

\section{results}

\subsection{Fornax morphology}
\begin{figure}
\includegraphics[width=8cm]{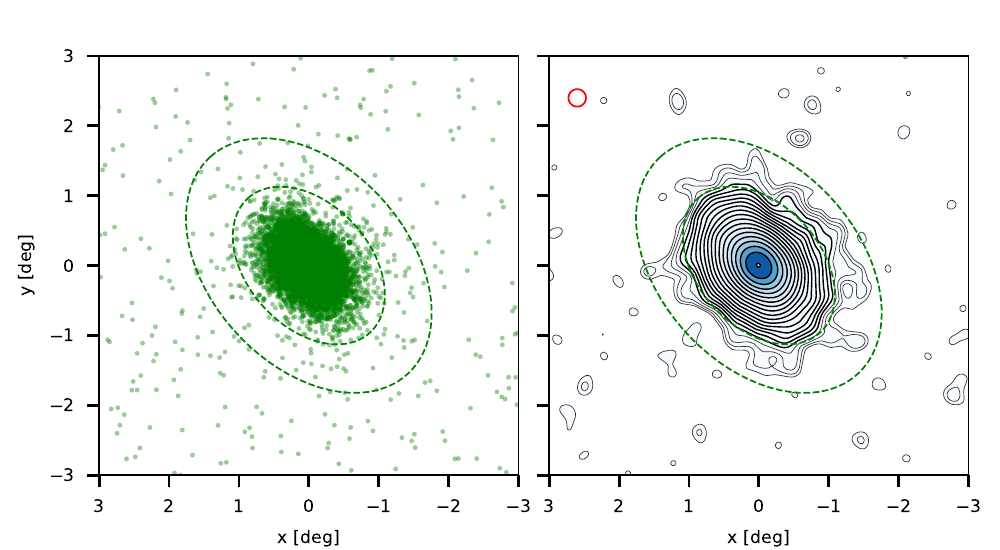}
\includegraphics[width=8cm]{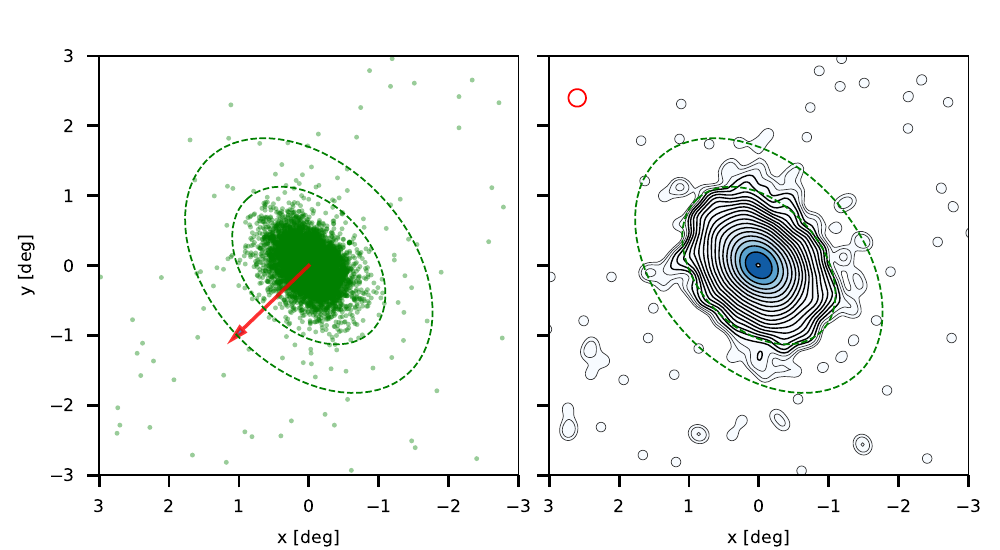}
\includegraphics[width=8cm]{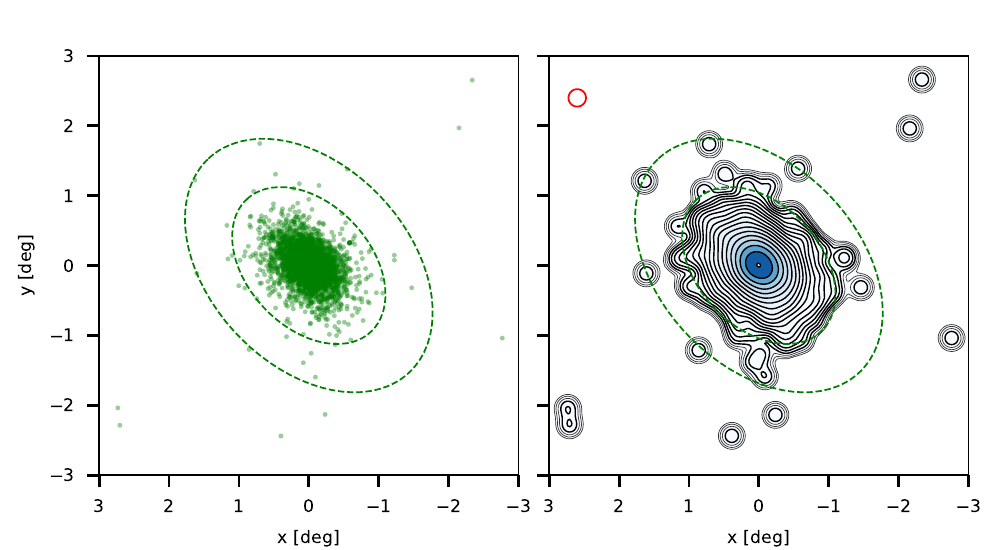}
    \caption{Morphology of Fornax by $S_0$, $S_1$ and $S_2$, from top to bottom, respectively. Direct dot plots are shown in the left panels, and the smoothed contour maps are shown on the right. In all panels, two ellipses (gray dashed lines) of 1.3 and 2.1 degree in radius, describing the morphological shape (Table.~\ref{tab:fornax}),  are superposed. The red arrow in left-middel panel indicates the mean PM of Fornax (see Table~\ref{tab:fornax}). For $S_0$, contour levels range from $5.30\times10^{-3}$ to 14.1 counts/arcmin$^2$ with logarithmic interval factor 1.483. For $S_1$, $2.75\times10^{-3}$ to 9.87 counts/arcmin$^2$, the interval factor 1.490. For $S_2$, $0.740\times10^{-3}$ to 3.10 counts/arcmin$^2$, the interval factor 1.510. Note that the peak density is lower than that of 1D density profile, because the contours has been smoothed by a constant Gaussian kernel of 0.25 degree, indicated by a red circle at the top-left corner. In all contour maps, thin and dark-gray contours are between 3 to 10 sigma significance above the 2D background fluctuations (see Table~\ref{tab:stat}). }
    \label{fig:morph}
\end{figure}

The spatial distribution of the three samples are shown in the left panels of Figure~\ref{fig:morph}. A smoothed contour map of the spatial distribution is shown in the right panels. We applied a fixed Gaussian kernel of FWHM 0.25 degree for smoothing. Only the central 6x6 sq degrees is shown; outside this scale no substructures can be identified.  We defined an annulus covering 6 to 10 degrees as the background reference, and calculated the mean densities and fluctuations for each sample (see Table~\ref{tab:stat}).
Morphologies from different samples show that the main body (those black contours above 10-sigma level) of Fornax appears quite regular and substantially symmetric. Overall, distributions of the three samples between 1.3 and 2.1 degrees are in good agreement, despite some variations. 
$S_0$ shows a possible extended feature on the minor axis in the north-west direction, at position $(x,y)=(-0.8,0.5)$, i.e., the last black contour. This feature is not confirmed in $S_1$ and $S_2$. Sample $S_2$ shows less significant extended features, though there are still few stars at 2.1 degrees, see also Figure~\ref{fig:cmdofpm}-$d$ and the relevant discussion in Sect.~\ref{sec:selCMD} about the significance.\\

\subsection{Surface Density Profile of Fornax}
\label{sec:radialprofile}
We derive and analyze the radial surface-density profile for each member candidate sample by applying the same algorithm as described below. For simplicity, we choose to present  $S_1$ in this section, while $S_0$ and $S_2$ are presented in Appendix~\ref{sec:dens02}. 

\begin{figure*}
	\includegraphics[width=15.cm]{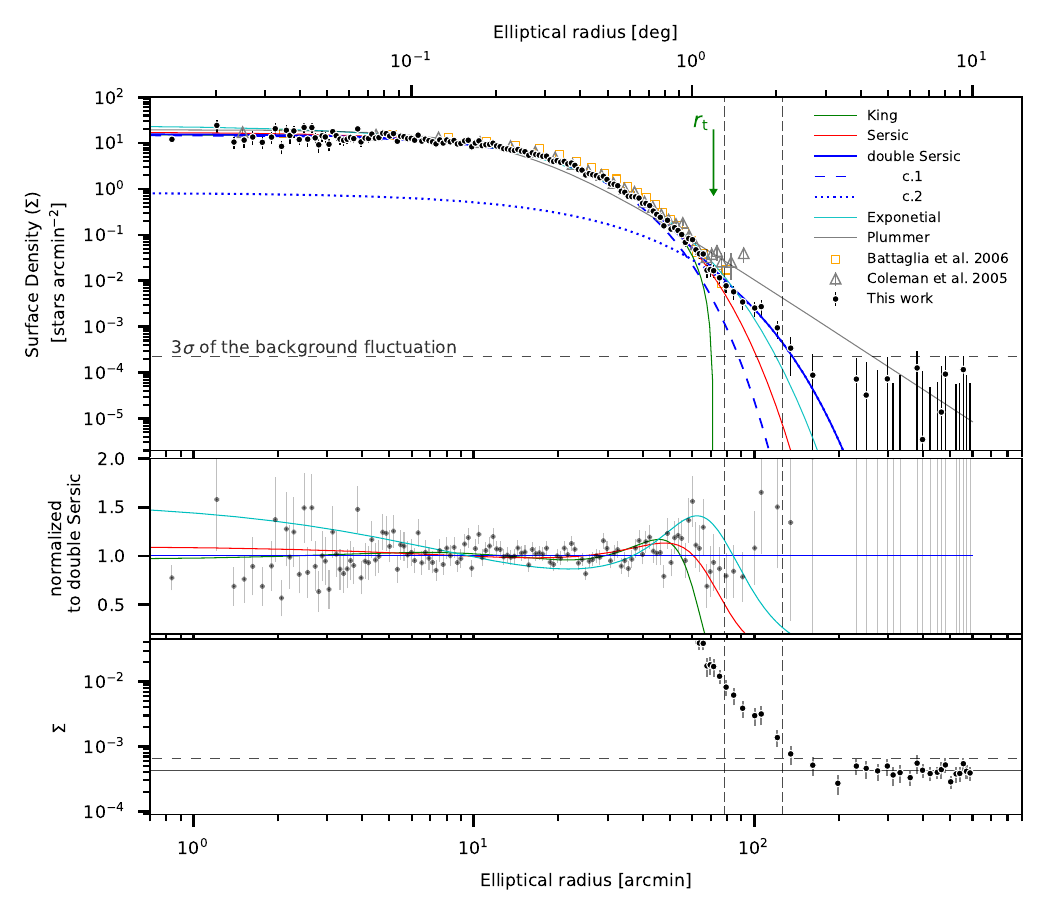}
    \caption{Surface density profile indexed by elliptical radius for $S_1$. In the main panel (top), the corresponding background densities have been subtracted. The horizontal dashed lines indicate the detection limits (Table~\ref{tab:stat}). Different best fitting theoretical density profiles and observed profiles are shown (see the legend). {The two components of the double-Sersic profile are denoted by "c.1" and "c.2".} The vertical dashed lines indicate 1.3 and 2.1 degree, respectively. 
    The green arrow indicate the location of the best-fitting King tidal radius $r_{\rm t}$.
    Two diagnostic panels are shown below: 1) Normalised data (dimmed into gray) and different models with respect to the double-Sersic model; 2) the density profile zoomed to the background level (gray solid and dashed lines for the mean and 3-sigma uncertainty, respectively), for inspecting how density profiles vary at large radii before subtracting the mean background. 
    }
    \label{fig:denprof}
\end{figure*}

Following the method of \citet[see also \citet{Battaglia2006}]{Irwin1995}, the radial profile is built directly from the candidate catalog, by evaluating the mean density in each elliptical annulus indexed by the semimajor axis. We assumed a mean geometry, i.e an P.A. and ellipticity as listed in Table~\ref{tab:fornax}, for each candidate sample, respectively. Such an assumption is helpful to explore the extent of a dSph in the low-density outskirts and can provide an easy way to understand and interpret the results, \citep[see][for a detailed discussion]{Irwin1995}.
Figure~\ref{fig:denprof} shows the radial surface-density profile for $S_1$.
Error bars are estimated from Poisson noise based on the counts in each annulus. In order to see details, we used high-resolution and logarithmic spacing in radius. We also request a minimal count of 9 stars in a single radius bin (resulting in a 33\%  statistic error at most), otherwise, it  will be combined into its neighboring annulus. Then the profile is obtained with adaptive resolution especially at center and outskirts.
We have examined the background uncertainty with two different methods. Method-1 calculates a mean number density within the background reference, i.e. the annulus from 6 to 10 degree, and the Poisson noise  of counts as its uncertainty.  Method-2 calculates the background density using the error weighted mean of all stars lying between 6 to 10 degree from the density profile \citep{Battaglia2006}. The latter include background fluctuations. Both methods provide the same mean value, and we have chosen the larger error bar as a conservative estimation of the background uncertainty. Under such considerations, for $S_2$, we have chosen the Poisson noise as the estimate of background uncertainty, while for $S_0$ and $S_1$, we have adopted the uncertainties using Method-2. As indicated in the figure, after subtracting the background, we are able to trace the density profile by almost 5 decades from the center to the outskirts above the 3-sigma level (defined as detection limit) of the background fluctuations (Table~\ref{tab:stat}).
 Our results reach 2 decades, i.e., 5 magnitudes, deeper than previous studies mentioned in Table~\ref{tab:method}.

Here we evaluate the depth or the equivalent surface brightness limit of our results. To do so, we first assume that our density profiles in the central region trace the same structure, i.e. density profile, as found by \citet{Munoz2018}. This underlying assumption is based on the fact that the RGB dominates the stellar light in Fornax, and are considered to be representative of the whole Fornax surface brightness made by \citet{Munoz2018}, who included other stellar population besides RGB stars. 
Then, we re-scale the density profile by matching the mean density inside the 0.3-degree elliptical radius to the effective surface brightness $24.77\pm0.16$ mag arcsec$^{-2}$ 
 in the V-band (defined inside the same elliptical radius) found by \citet{Munoz2018}.  The mean density inside the 0.3-degree elliptical radius are 17.318, 11.901, 3.612 star/arcmin$^2$ for $S_0$, $S_1$ and $S_2$ respectively. These values are applicable to the calibration of the corresponding 2D density maps. 
The reached surface brightness limits are listed in Table~\ref{tab:stat}. The statistical error of evaluating the mean density is very small because of the large number of counts, so the uncertainty of the equivalent surface brightness is mostly coming from the physical calibration by \citet{Munoz2018}, i.e., 0.16 mag.

\begin{table} 
\caption{Best-fitting parameters of different theoretical density profiles : 
King model (core radius, $r_{\rm c}$, tidal radius, $r_{\rm t}$), 
Sersic model (Sersic radius, $R_{\rm S}$, shape index, $m$), 
exponential model (scale radius, $r_{\rm s}$) 
and Plummer model (scale radius, $b$). Peak density of each profile $I_0$ are all in unit stars/arcmin$^2$. The last columns give the reduced and the mean $\chi^2$, respectively. The latter is evaluated for the points between 1.3 to 2.1 degree. The goal is to compare between Sersic and double-Sersic model. }
\small
\label{tab:fit}
\centering
\begin{tabular}{llccccc}
\hline
\hline
Model$\!\!\!\!$  & \multicolumn{2}{l}{{ Sample$\!\!\!\!$$\!\!\!\!$}} &   \multicolumn{2}{l}{{parameters}}   & $\!\!\!\!$  $\chi^2/\nu$ & $\!\!\!\!$$\overline{\chi^2_*}$ \\ \hline
\multicolumn{2}{l}{King}  & $I_{\rm 0,K}$  &  $r_{\rm c}$ [$^\prime$] &  $r_{\rm t}$ [$^\prime$] &  \\
     & $S_0$ &  33.9$\pm$3.6 &  15.3$\pm$1.7     & 76.0$\pm$5.2    & 1.16 & - \\
     & $S_1$ &  24.4$\pm$2.8 &  15.4$\pm$1.9     & 71.6$\pm$5.2    & 1.28 & -   \\
     & $S_2$ &  ~7.5$\pm$1.3 &  14.7$\pm$2.5     & 76.3$\pm$8.7    & 0.88  & - \\ \\
\multicolumn{2}{l}{Sersic}  &  $I_{\rm 0,S}$ &  $R_{\rm S}$ [$^\prime$] &  m &  \\
     & $S_0$  &  25.4$\pm$2.7 &  14.5$\pm$0.6    & 0.83$\pm$0.04    & 1.41 & 5.15\\ 
     & $S_1$  &  17.4$\pm$2.0 &  14.7$\pm$0.7    & 0.80$\pm$0.04    & 1.26  & 4.64 \\ 
     & $S_2$  &  ~5.8$\pm$1.0 &  14.1$\pm$1.0    & 0.84$\pm$0.07    & 0.92 & 3.83\\ \\
\multicolumn{2}{l}{double-Sersic}   &  & & &  \\
  \multicolumn{1}{r}{ c.1}  & \multirow{2}*{$S_0$} &   22.3$\pm$2.6 &  15.8$\pm$0.8     & 0.74$\pm$0.05   &  \multirow{2}*{ 1.07 }  & \multirow{2}*{0.94}  \\
  \multicolumn{1}{r}{ c.2}  &  &   ~1.0$\pm$0.8 &  18.7$^{+3.3}_{-6.5}$     & 1.00$^{+0.12}_{-0.29}$    &  \\ \\

  \multicolumn{1}{r}{ c.1}  & \multirow{2}*{$S_1$} &   15.0$\pm$1.9 &  15.9$\pm$0.9     & 0.71$\pm$0.05   & \multirow{2}*{0.97} & \multirow{2}*{0.60}\\
  \multicolumn{1}{r}{ c.2}  &  &   ~0.83$\pm$0.65 &  19.7$^{+3.3}_{-6.6}$     & 0.91$^{+0.11}_{-0.29}$    &  \\ \\

  \multicolumn{1}{r}{ c.1}  & \multirow{2}*{$S_2$} &   ~4.8$\pm$0.9 &  15.5$\pm$1.2     & 0.73$\pm$0.08   & \multirow{2}*{0.78} & \multirow{2}*{0.37}\\
  \multicolumn{1}{r}{ c.2}  &  &   ~0.33$\pm$0.3 &  19.6$^{+4.0}_{-12.}$     & 0.92$^{+0.14}_{-0.90}$    &  \\ \\
\multicolumn{2}{l}{Exponetial} &  $I_{\rm 0,E}$  &  $r_{\rm e}$ [$^\prime$] &  &  \\
     & $S_0$   &  34.5$\pm$3.6 & 10.7$\pm$0.4     &     & 2.48  & - \\ 
     & $S_1$   &  24.6$\pm$2.8 & 10.3$\pm$0.5     &     & 2.47  & - \\ 
     & $S_2$   &  ~7.65$\pm$1.3 & 10.6$\pm$0.7     &     & 1.25  & - \\ \\
\multicolumn{2}{l}{Plummer}  &  $I_{\rm 0,P}$ &  $b$ [$^\prime$] &  &  \\
     & $S_0$   &  27.0$\pm$2.8 & 16.4$\pm$0.7     &     & 10.85  & - \\
     & $S_1$   &  19.6$\pm$2.2 & 15.4$\pm$0.7     &     & 12.01  & - \\
     & $S_2$   &  6.1$\pm$1.1 & 16.0$\pm$1.1     &     & 4.56  & - \\
\hline
\end{tabular}
\end{table}

Following \citet{Battaglia2006}, we fit different theoretical density profiles to the data, and the results are listed in Table~\ref{tab:fit}.  For all theoretical profiles we have adopted the same notation as \citet{Battaglia2006}. First, let us compare the results between the three samples. For each theoretical profile, the best fitting parameters are similar from one sample to another, within the error bars. We also note that the reduced-$\chi^2$ generally decreases from $S_0$ to $S_2$, which could reflect the increase of the fraction of true Fornax members from  $S_0$ to $S_2$, but could also possibly be due to the decrease of the sample size which leads to less statistical significance.

Second, let us focus on $S_1$. Fornax extends much farther than the King tidal radius (see Table~\ref{tab:fit}), outside which the observed density profile shows a change of slope, in a similar way as the observed breaks found in other classical dSphs \citep[e.g.,][]{Westfall2006}.
Furthermore, we can safely reject the Plummer profile, according its huge reduced-$\chi^2$ value. 
Although the exponential profile gives a reasonable description of the data, it shows some systematic biases at different radii. This can be seen more clearly in the middle panel of Figure~\ref{fig:denprof} where the best-fit of exponential profile deviates from the data points.

The Sersic profile provides a very good fit to the data out to 1~degree (see, the lower panel in Figure~\ref{fig:denprof}). The departure (i.e., the break) beyond this radius reveals that an additional component exists in Fornax. To characterize this additional component we introduce a double-Sersic model to decompose the density profile. It gives an obvious improvement of the fitting, as shown by the significant changes of the mean $\chi^2$ indicated in the last columns of Table.~\ref{tab:fit}, suggesting the need of this second component.  The second Sersic component , i.e., "c.2" in Figure~\ref{fig:denprof}, starts to dominate the density at 0.9 degree, where the two components reach the same density.  It includes 10, 9, and 12.4\% of the total net counts by integrating the theoretical profiles for $S_1$, $S_0$, and $S_2$ respectively.  Accounting for the 2D morphology that shows a nearly symmetric morphology, it suggests that the second component behaves like a surrounding halo, though there could be projection effects.

We find that within 1 degree, the density profiles derived for Fornax in this paper are almost the same as the result obtained by \citet{Coleman2005} (see Figure~\ref{fig:denprof} open triangles). Interestingly, \citet{Coleman2005} used only bright RGB star selection too, covering an even shallower 2.5 mag of RGB stars. 
 We have re-scaled their profile by normalizing the average density within 0.1 degree.   
The case "All" in \citet{Battaglia2006} has also been re-scaled and plotted in Figure~\ref{fig:denprof}. There is a visible discrepancy between 0.3 to 1 degree, which agrees with fitting results. For example, with a single Sersic profile and $S_1$, we found a smaller characteristic radius than \citet{Battaglia2006}, i.e.,14$^\prime$.7 versus 17$^\prime$.6. We note, however, that our result is consistent with the result by \citet{Bate2015} after their correction of the crowding effect in the center of the galaxy.  These comparisons are in line with what we discussed and expected in Sect.~\ref{sec:method}, the relatively more shallow photometric data of Gaia and its good image quality could lead to results much less affected by the crowding effect at the center of Fornax.

\section{Discussion and Conclusion}
Thanks to Gaia's homogeneous data quality and full coverage of PM measurements, we have studied the stellar structure of Fornax dSphs over a huge area (400 sq deg), and reached an extreme depth of photometry, down to 12 magnitudes fainter than the central density of  Fornax, which is equivalent to a surface brightness $37.94\pm0.16$~mag/arcsec$^2$  in the V-band.  Our study demonstrates that using Gaia's multiple and independent parameter spaces, i.e. CMD, PM, and parallax, allows to perform a member candidate selection that provides a breakthrough for studying dSph outer structures compared to former studies with ground-based deep observations.
Above the detection limit, we have discovered a significant second component in the Fornax dSph, which resembles a stellar halo due to its nearly symmetric morphology. This component represents about 10\% of the total mass of Fornax, and extends out to 2.1 degree (i.e. 5.4 kpc or almost 7 times the half-light radius). This additional component in the Fornax outskirts is well described by a Sersic profile with index of $0.92^{+0.11}_{-0.29}$, which is consistent with an exponential profile. This might suggest that Fornax could extend even further.  What is the origin of this halo-like structure? 

\citet{Battaglia2015} presented a set of dedicated simulations to fit Fornax, in a frame for which it is DM dominated and a long-lived satellite of the MW. All their models but one (assuming that mass follows light), can reproduce quantitatively both the $\sigma_{\textsc{los}}$ radial profile and the observed morphology out to 1.3 degree \citep{Battaglia2006}. In these models the stellar component of the simulated object is very stable against tidal stripping that removes only up to 10\% of the initial stellar mass after 12 Gyr of orbital integration. However, none of these models predict a second component similar to that we found from Gaia data. 
Using cosmological simulations, \citet{Genina2020} study analogs of Fornax according to its star formation history. They argue that the Fornax dSph may show tidal tails at around 35-36 mag/arcsec$^2$, which is only 6 magnitudes in contrast to the central density of the simulated object (see their Fig.~6). With our Gaia results, we have been able to trace the density profile of Fornax within a 12 magnitude range, and we do not find any tidal tail over a large field surrounding Fornax (up to 400 sq degrees). 
Conversely to clues of tidal stripping in Fornax, we found that the outskirt of Fornax, i.e. $r_{\rm ell} > 1.3$ degree, resembles a halo-like structure because it appears rather symmetric. It suggests that other mechanisms than tidal stripping are at work in Fornax. Among them, a former dwarf-dwarf merger was supported by the discovery of a shell structure near Fornax \citep{Coleman2004}, but has been disproved later by \citet{Bate2015}.\\ \\

Alternatively, the MW dSphs could be affected by tidal shocks \citep{Hammer2018,Hammer2019}, and in the following we investigate whether it could naturally explain the secondary component of stars in the Fornax outskirts. The mechanism includes the transformation of a Fornax progenitor that was a gas-rich dwarf galaxy.  Fornax shows very recent star formation (50-100 Myr ago, \citealt{Coleman2008}), and if we were observing this galaxy at that epoch it would have resembled a gas-rich dwarf irregular (dIrr, \citealt{Battaglia2013}). The case of a DM-free progenitor was first investigated by \citet{Yang2014} with numerical simulations. If gas was dominating the former Fornax dwarf, after the last gas removal, the residual stars would have expanded following a spherical geometry due the loss of gravity. As shown by \citet[and references therein]{Hammer2019}, the tidal stripping term becomes negligible and energy exchanges with the MW are dominated by tidal shocks, which can dominate the \textsc{los}-kinematics. Not all stars are in resonance with the MW or tidally shocked, an average fraction of only 25\% is sufficient to explain the observed hot kinematics. We suggest that the second stellar component revealed by Gaia data may be due to the recent expansion of stars in Fornax. Detailed numerical modeling will be required to reproduce the overall properties of Fornax in the tidal shock scenario.

Such a scenario assumes that the MW halo is filled by diffuse and ionized gas (up to few million Kelvin), know as the Circumgalactic Medium \citep[CGM,][]{Tumlinson2017}. There is direct evidence of such diffuse gas surrounding the MW \citep{Miller2013,Anderson2010}. There are also many indirect indications for the existence of this diffuse gas, such as the strong dichotomy of the distribution of dSphs and dwarf irregular galaxies within and beyond 300 kpc, respectively \citep{Grcevich2009}.  The presence of head-tail High-Velocity Clouds (HVCs) \citep{Kalberla2006} at $>$ 50 kpc from the Magellanic Stream implies a density of ionized gas of about $10^{-4} {\rm cm}^{-3}$ at 50-70 kpc and $10^{-5} {\rm cm}^{-3}$ out to a few hundred kpc. When gas-rich dwarf galaxies fall into the MW, their cold gas will feel the ram pressure generated by the hot gas in the MW halo, and later on they could be fully stripped. The best example for such an on-going process is The Magellanic Stream, since the ram pressure tails lagging behind the Magellanic Clouds are indeed observed and reproduced by a 'ram-pressure+collision' model \citep{Hammer2015,Wang2019}. It is reasonable that such a strong ram pressure force could remove the gas from the in-falling dwarf galaxies \citep{Grcevich2009,Yang2014}. Fornax can be interpreted as an archetype for this scenario, because of its very recent star formation implying a recent removal of its gas. \citet{Rusakov2021}, after examining possible scenarios and simulations, found that the observed star formation history (SFH) of Fornax is better explained by periodic passages about the MW. But their proposition cannot explain the two star formation peaks within the last Gyr (see their Fig.~12), because the last orbital period of Fornax is at least 2 Gyr \citep{Rusakov2021}. 
Our scenario could explain naturally the subsequent star formation history of Fornax. After Fornax's progenitor (which in this scenario is supposed to be a gas-rich and dark matter free dwarf galaxy) entered the MW halo, around 2 Gyr ago, the gas in the dwarf galaxy will be compressed due to ram pressure, igniting star formation. Because there is no dark-matter, which means the potential of the galaxy is very shallow, the feedback of star formation could push the gas to larger radii of the galaxy, hence reducing or even halting the star formation process.  As gas is cooling down and falls back to the center of the galaxy, with the help of ram pressure a new cycle (i.e. epoch) of star formation will be activated, and so on. This process could explain the multiple epochs of star formation in the last 1 Gyr.  The oscillation of such a star formation process depends on the mass of the dwarf galaxy and the strength of ram pressure, and thus could put some constraints on the mass of Fornax.
As Fornax approached its current position, it experienced the last ram-pressure process, and then lost its last gas in the recent few hundred Myr. The last event of very recent gas removal is consistent with the observed 100-300 Myr stellar population in Fornax.
If confirmed, the HI cloud superposed on Fornax and discovered by \citet{Bouchard2006} could be the last gas cloud that is leaving the Fornax dSph now.

Our discovery of this second component, i.e. a halo-like stellar component in Fornax, motivates us to revisit the understanding of the "breaks" in dSphs density profiles. 
With Fornax, almost all dSphs, except Leo~II, 
are reported to display such a break in their density profiles. If this second component of Fornax can be explained as the result of expanding stars, how about the other dSphs? It could be interesting to identify the morphologies of the second component responsible of the break in their density profile, and whether they are consistent with tidal stripping or alternatively with tidal shocking. The latter could be verified with internal kinematics via PMs by Gaia in the future, see our Introduction or \citet[][]{Hammer2018,Hammer2019}.

\section*{Acknowledgements}
The authors would like to thank Fr\'ed\'eric Arenou, Carine Babusiaux, Piercarlo Bonifacio, Jianling Wang, and Hefan Li for their helpful disucssions. 
MSP acknowledges funding of a Leibniz-Junior Research Group (project number J94/2020) via the Leibniz Competition, and thanks the Klaus Tschira Stiftung and German Scholars Organization for support via a KT Boost Fund.
This work has been supported by the National Natural Foundation of China (NSFC No. 12041302 and No. 11973042). We also thank for its support the International Research Program Tianguan, which is an agreement between the CNRS, NAOC and the Yunnan University.
This work presents results from the European Space Agency (ESA) space mission Gaia. Gaia data are being processed by the Gaia Data Processing and Analysis Consortium (DPAC). Funding for the DPAC is provided by national institutions, in particular the institutions participating in the Gaia MultiLateral Agreement (MLA). The Gaia mission website is https://www.cosmos.esa.int/gaia. The Gaia archive website is https://archives.esac.esa.int/gaia.

\section*{Data Availability}
The data underlying this article are available at https://doi.org/10.5281/zenodo.5021235.
We provide the catalog of Fornax member candidate, including the three final samples, as well as the data of surface density profiles. 

\section*{}{\bf Note added in proof:} 
During the publication of the paper, we noticed
that stellar halos of MW dSphs were also discovered recently in the
Fornax \citep{Stringer2021}, and Tucana II \citep{Chiti2021} sourroundings. These findings confirm our result and motivate us to
further study other dSphs.







\appendix
\section{Density profiles for S0 and S2 }
\label{sec:dens02}
\begin{figure*}
	\includegraphics[width=12.cm]{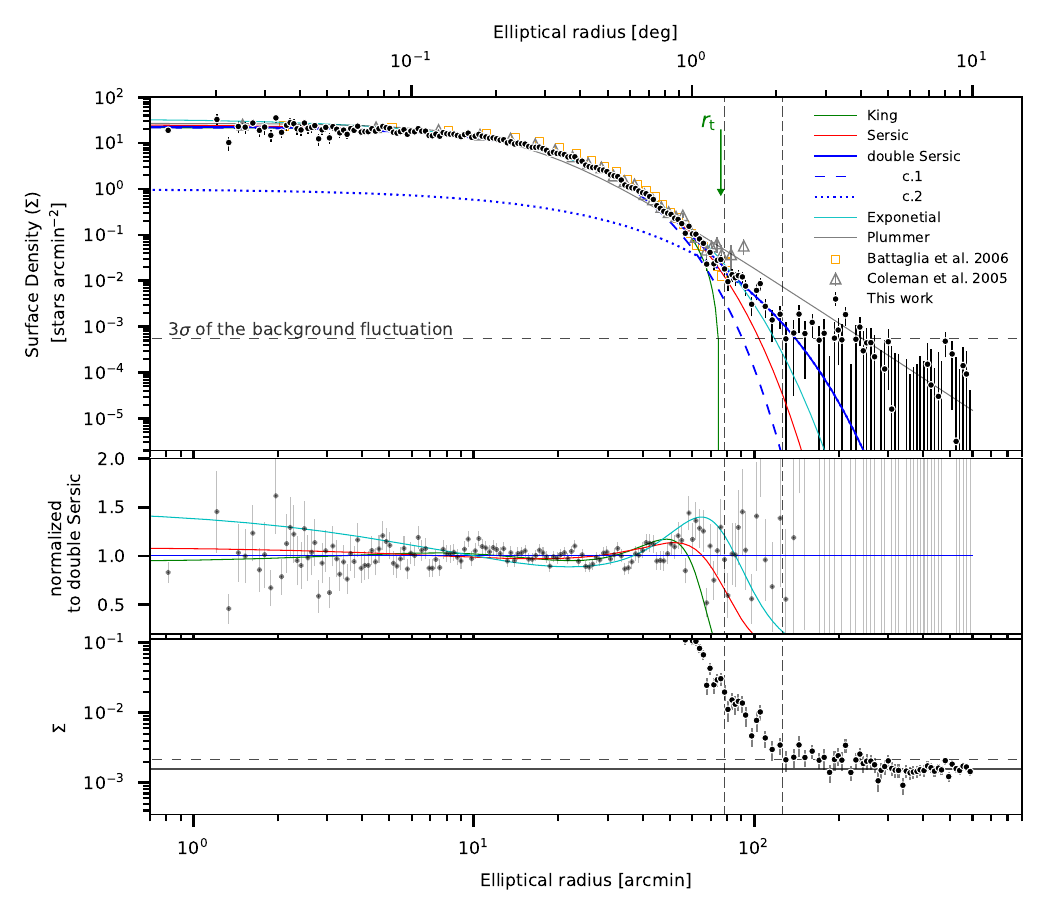}
	\includegraphics[width=12.cm]{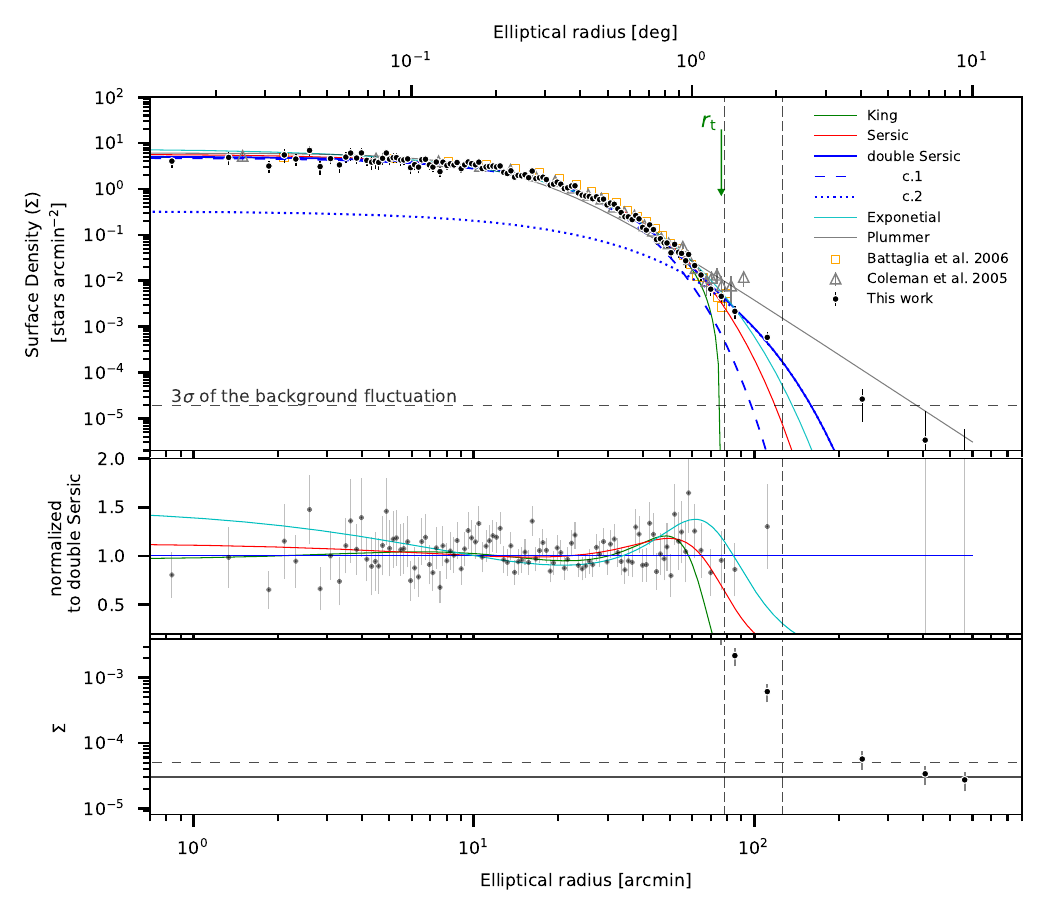}
    \caption{Surface density profiles for $S_0$ (top) and $S_2$ (bottom), see Figure~\ref{fig:denprof} for a detailed description. Note that, for $S_0$, there is a hint of over density in between 2.1 and 5 degree when comparing to our fiducial background region (6-10 deg). We suspect this is mostly due to the contamination of QSO because this over-density is disappeared when we switch to $S_1$ and $S_2$ by limiting the color \texttt{bp\_rp} to be redder, also because we have identified that this weak over-density (total S/N~$=1.8$) is mostly caused by the sources located around $(x,y)=(-2,2)$ in $S_0$ sample. Color contamination due to QSO is shown in Figure~\ref{fig:cmdselections}. 
    }
    \label{fig:denprof2}
\end{figure*}

Figure~\ref{fig:denprof2} shows the 1D surface density profiles derived from $S_0$ and $S_2$. 

\section{Density profile by Irwin \& Hatzidimitriou 1995}
\label{sec:ih95}
\begin{figure*}
	\includegraphics[width=6cm]{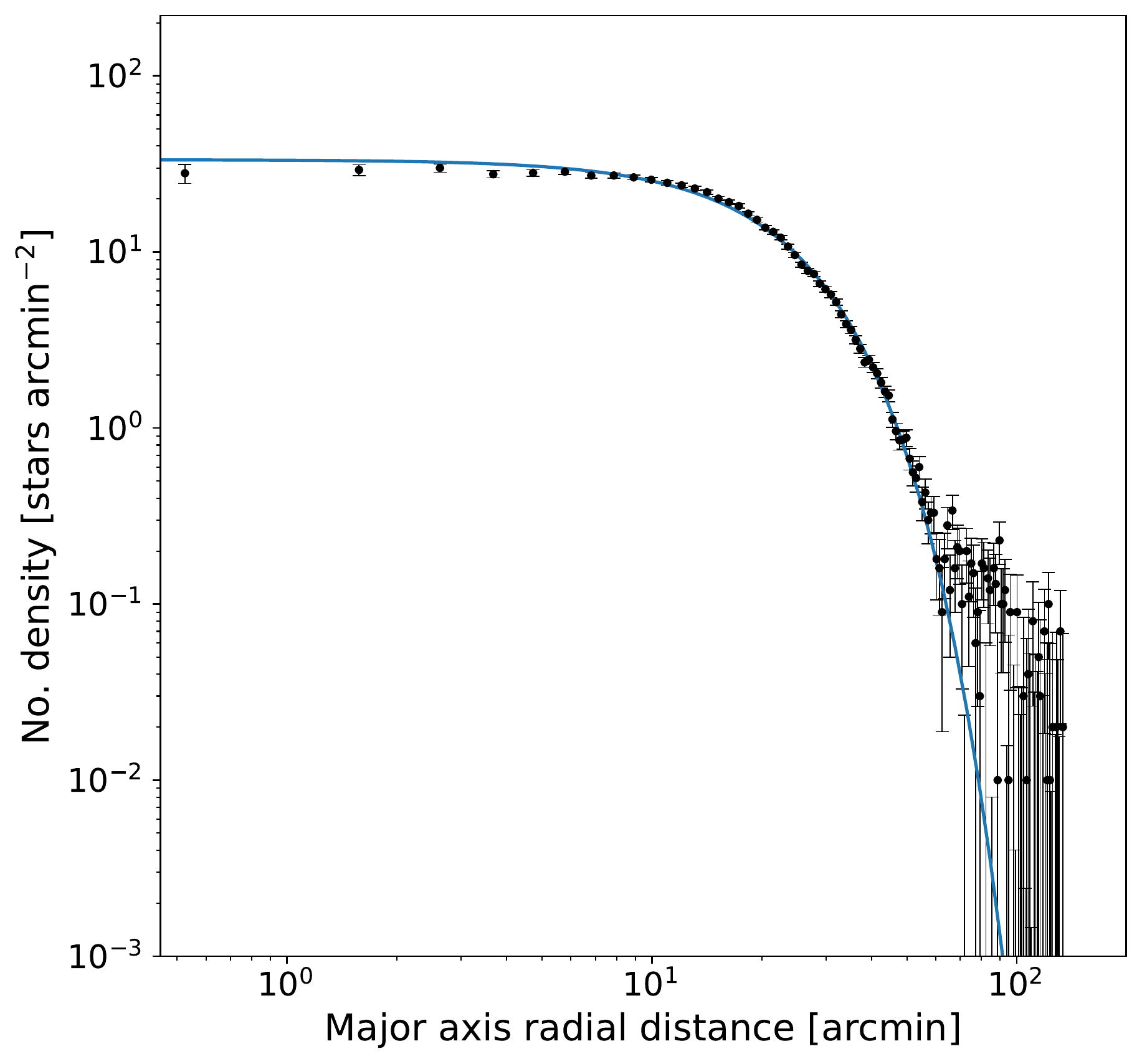}
    \caption{Using the same method described in Sect.~\ref{sec:radialprofile} we fit a single Sersic profile to the radial profile of Fornax derived by \citet[][see their Table 3]{Irwin1995}. We find a Sersic radius $R_S = 21.88 \pm 0.64$ arcmin.}
    \label{fig:ih95}
\end{figure*}

Figure~\ref{fig:ih95} shows the density profile derived by \citet[][]{Irwin1995} and our fit of the Sersic profile.

%


\bsp	
\label{lastpage}
\end{document}